\documentclass{ws-ijqi}

\newcommand{\ket}[1]{|#1\rangle}
\newcommand{\bra}[1]{\langle#1|}

\newcommand{\bk}[2]{\langle#1|#2\rangle}
\newcommand{\eq}[1]{Eq.~(\ref{#1})}
\newcommand{\eqs}[1]{Eqs.~(\ref{#1})}
\newcommand{\fig}[1]{Fig.~(\ref{#1})}

\begin{document}

\markboth{S. M. Giampaolo, F. Illuminati, A. Di Lisi, and G.
Mazzarella} {Massive Quantum Memories by Periodically Inverted
Dynamics}

%
\catchline{}{}{}{}{}
%

\title{MASSIVE QUANTUM MEMORIES BY PERIODICALLY \\
       INVERTED DYNAMIC EVOLUTIONS}

\author{S. M. Giampaolo, F. Illuminati, A. Di Lisi, and G. Mazzarella}

\address{Dipartimento di Fisica ``E. R. Caianiello'',
Universit\`a di Salerno; \\CNR-Coherentia, Gruppo di Salerno; and
INFN Sezione di Napoli, \\Gruppo Collegato di Salerno, Via S.
Allende,
84081 Baronissi (SA), Italy.\\
giampaolo@sa.infn.it; illuminati@sa.infn.it; dilisi@sa.infn.it;
mazzarella@sa.infn.it}

\maketitle

\begin{history}
\received{(23 June 2005)}
\end{history}

\begin{abstract}
We introduce a general scheme to realize perfect quantum state
reconstruction and storage in systems of interacting qubits.
This novel approach is based on the idea of controlling the
residual interactions by suitable external controls that,
acting on the inter-qubit couplings, yield time-periodic
inversions in the dynamical evolution, thus cancelling exactly
the effects of quantum state diffusion. We illustrate the
method for spin systems on closed rings with $XY$ residual
interactions, showing that it enables the massive storage of
arbitrarily large numbers of local states, and we
demonstrate its robustness against several realistic
sources of noise and imperfections.
\end{abstract}

\keywords{Quantum Information; Quantum Control; Spin Systems.}

\section{Introduction}

Along the pathway that should eventually lead to the realization
of scalable schemes of quantum computation, much theoretical work
has been recently devoted to develop physical devices
for the efficient processing and the coherent transfer of
quantum data\cite{Bose}$^-$\cite{Datta}.
Besides these two fundamental aspects, a further crucial
requirement needed for the realization of a quantum computer is the
possibility to store quantum information on a time scale at least
comparable to the one needed to perform computational tasks.
In particular, it would be very important
to introduce systems acting as stable and robust quantum memories
that recover and conserve large sets of quantum states that would be
otherwise usually lost in very short times, due to quantum diffusion
and decoherence\cite{Song}.

An ideal quantum register is formed by a set of non interacting
identical two--level systems (qubits), in the sense that either
the Hamiltonian describing the dynamics of the internal degrees of freedom
of the qubits, as well as the Hamiltonian describing the inter-qubit
interactions, can be changed at will to realize the desired
quantum-state manipulations. However, when we move from the ideal
case to the arena of possible concrete realizations, the qubits cannot
in general be considered identical due to the unavoidable presence of
local imperfections, and moreover, residual inter-qubit couplings
are always present. The existence of these two sources of noise
causes, as a consequence, the loss of information in the quantum register.

To ensure the ability to efficiently store quantum data, many
different noise-evading schemes have been
proposed\cite{Sun}$^-$\cite{Kitaev}
All these works can be classified in two different groups: either
the proposed schemes are based on some error correction technique;
or they rely on some intrinsic mechanism, holding at least for a
specific subset of quantum states, that leaves the information
unaffected during the time-evolution of the quantum register
(decoherence-free subspace schemes). The second approach is the
most desirable one as it provides in principle a radical
solution to the storing problem but, unfortunately, it is very
sensitive to almost any source of imperfection. On the contrary,
the schemes based on quantum error correction techniques are
characterized by a dynamics that allows at any time the
unambiguous reconstruction of the initially stored state, at least
for a subset of states. The main trouble with error correction
techniques lies in the fact
that usually only very few states can be effectively accessed to
store information and, hence, relatively large arrays of qubits
are needed to store relatively small amounts of information.

In the present work, starting from Section \ref{general},
we discuss and generalize a new approach to quantum state storage
that has been recently introduced\cite{precedente}.
The basic idea of this novel scheme for the realization of massive
quantum memories is to exploit
time-modulated periodic dynamics that allow a
perfect, periodic reconstruction of a generic initial state.
In some sense, this new method can be seen as a time-periodic
generalization of the decoherence-free subspace approach.
We will show that our scheme is very robust against several
sources of noise and imperfections.

An explicit realization of this approach is illustrated in Section
\ref{AB}, in which a topological linked magnetic flux with
periodic step-time modulation is used to cancel the
effects of the residual $XY$ interactions with site-dependent
couplings induced by local imperfections. Finally,
in Section \ref{conclusion} we summarize our results and discuss
future perspectives.

\section{Storing scheme: overview and prescriptions}
\label{general}

As we have anticipated, a quantum register can be seen as an
ensemble of qubits each one of them described by its own internal
Hamiltonian. In the ideal case the internal interactions of the
qubits are all identical and residual interactions between them
can be switch on and off at will. However, realistic realizations
of arrays of qubits strongly deviate from the ideal case: internal
interactions are different due to imperfections, and residual
inter-qubits interactions are always present. For this reason,
realistic quantum processors and registers are in general properly
described by generic Hamiltonians of the
form\cite{Georgeot1}$^-$\cite{Montangero}
\begin{equation}\label{Generic Hamiltonian}
    \hat{H} \, = \, \hat{H}_0 + \hat{H}_{err} \; ,
\end{equation}
where $\hat{H}_0$ stands for the Hamiltonian of ideal model
register and $\hat{H}_{err}$ for the undesired noise terms. Due to
the presence of the noise terms, immediately after having stored a
quantum state in the register (in the following we always refer to
it as the {\em initial state} $\ket{\psi_0}$), such state starts
to evolve, and hence the fidelity of the quantum register
decreases, progressively corrupting the storing process. To
overcome this problem we introduce a scheme that is based on the
idea of modulating the evolution generated by the noise terms,
rather than trying to completely eliminate it. In fact, we suggest
to engineer the quantum register, and the corresponding quantum
information storing process, in such a way that the overall
dynamics yields a perfect (or almost perfect) periodic
reconstruction of the initially stored state, with an a priori
determined period $T$.

The idea to obtain a periodic state reconstruction, naturally
leads to consider time-dependent, periodic Hamiltonians. Our
approach aims in fact at modulating one or both of the two
contributions in \eq{Generic Hamiltonian} to obtain the
time-dependent periodic Hamiltonian that realizes the desired
state reconstruction. As it is well known, in the presence of a
generic time-dependent Hamiltonian, the quantum state evolution
can be tracked down by resorting to the Dyson series
representation. Obviously, the structural complexity of the Dyson
series does not allow to identify all the possible Hamiltonian
dynamical evolutions yielding perfect time-periodic state
reconstruction. However, the Dyson series can be easily resummed
when the Hamiltonian enjoys the property to commute at different
times: $[\hat{H}(t),\hat{H}(t')]=0$ for all $t$ and $t'\neq t$.
Thus, our first prescription is to conceive a quantum register
described by a time-periodic modulated Hamiltonian, commuting at
all times. In the case in which this property holds only in a
subspace of the whole Hilbert space of the quantum register, our
approach is still valid if the state to be stored is chosen in
this subspace.

Under the time-commutation hypothesis, we can determine a complete
set of states $\{ \ket{\alpha} \}$ that are eigenstates of the
time-dependent Hamiltonian at all times even if the correspondinge
energy eigenvalues are time-dependent functions
$\varepsilon^\alpha(t)$. Because the energy eigenstates form a
complete basis set, the initial state $\ket{\psi_0}$ can be
written as a linear combination: $\ket{\psi_0} = \sum_\alpha
c_\alpha \ket{\alpha}$, with $c_\alpha$=$\bk{\alpha}{\psi_0}$, and
the sum runs over the complete set of eigenstates. It is then
simple to write the evolution of the initial state after a time
$t>0$
\begin{equation}
\label{evoluzione} \ket{\psi (t)} \, = \, \sum_\alpha c_\alpha\exp
\left(-i\int_0^t \varepsilon^\alpha(\tau) d\tau
\right)\ket{\alpha} \; .
\end{equation}
From \eq{evoluzione} it is immediate to see that if at a certain
time $T>0$ all the integrals in the sum are equal or differ from
each other by integer multiples of $2 \pi$, then the initial state
$\ket{\psi_0}$ is perfectly reconstructed, but for an irrelevant
global phase factor. This is then the second requirement that one
needs to impose on the time-modulated dynamics in order to realize perfect
time-periodic quantum state storage.

In the following of the paper we will show a simple but
interesting implementation of this storing scheme based on the
Aharonov-Bohm effect in systems of charged particles on a lattice.

\section{Quantum register based on the Aharonov-Bohm effect}
\label{AB}

In this section we illustrate a specific realization of the
general approach discussed above, that enables to eliminate the
effect of the residual interactions between qubits described by $XY$
Hamiltonians that, as we will show in subsection \ref{nmod} rapidly
destroys the storing capacity of a quantum register. As it is well
known, spin Hamiltonians can be used as an effective description
of systems of hopping and/or interacting quantum
particles on a lattice, in the presence of an energy gap such that
only two local states on each lattice site can be
considered\cite{Schoen}. In general, in any such situation, an
effective two-level system can always be mapped in a formal
spin-$1/2$ one\cite{Feynman57}. If the particles are charged,
in the presence of a linked magnetic flux the real-valued hopping
amplitude between particles, or, in spin language, the nearest-neighbor
coupling, is transformed in a complex-valued
quantity\cite{Peierls}$^-$\cite{Scalapino}. In the presence of a
simple geometry such as a tiny solenoid placed in the center of a
circular ring of regularly spaced sites, the phase $\theta$ of the
nearest-neighbor interaction amplitude is proportional to the
magnetic flux: $\phi$ $(\theta \propto \phi/N)$ where $N$ is the
total number of qubits in the ring\cite{Peierls,Scalapino}. As
shown in subsection \ref{mod}, the magnetic flux can be suitably
modulated in time to perfectly eliminate the dephasing effects due
to the undesired residual interaction terms. Moreover, in
subsection \ref{err} we show that the storing scheme is robust
with respect to other possible sources of noise that can arise in
practical realizations.

\subsection{The unmodulated quantum register}
\label{nmod}

\begin{figure}[t!]
\includegraphics[width=7.5cm]{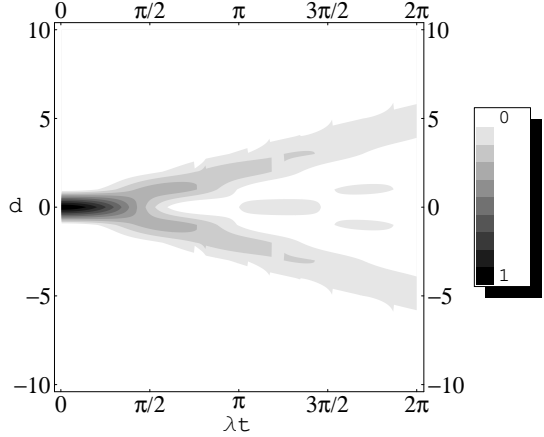}
\caption{Homogeneous coupling amplitude: $\chi_i= 0$. Contour plot
showing the evolution of the overlap ${\cal{F}}_{d}(t)$
as a function of the distance $d$
from site $0$ ($y$-axis) and of the dimensionless time $\lambda
t$ in atomic units $\hbar=1$ ($x$-axis). We choose $B=100
\lambda$. The value of the overlap increase from white
$({\cal{F}}_{d}(t)=0)$ to black $({\cal{F}}_{d}(t)=1)$.}
\label{figura1}
\end{figure}

We begin by considering a quantum register in which $N$ identical
qubits are placed on the sites of a closed ring, and, due to the
presence of local imperfections, are coupled by
nearest-neighbor $XY$ interactions with site-dependent amplitudes.
\begin{equation}\label{Hamiltonian 1}
\hat{H} \, = \, -\sum_{i}(\lambda +\chi_i)\bigg(
\sigma_{i}^{+}\sigma_{i+1}^{-} + H. c. \bigg) \, + \,
B\sum_{i}\sigma_{i}^{z} \; ,
\end{equation}
where $B$ is the half-energy gap between the two levels of the
qubits, $\lambda$ is the global, averaged nearest-neighbor
coupling amplitude, and $\chi_i$ is a site-dependent random
variable of vanishing mean, representing the local imperfection in
the coupling at site $i$. The presence of the coupling, even if
$\lambda \ll B$ and $\chi_i =0$ $\forall \; i$, rapidly destroys
the storing capacity of the register. To illustrate this effect,
let us introduce the following set of states
\mbox{$\ket{\Psi_d}=\ket{\uparrow_{d}}\prod_{i\neq d
}\ket{\downarrow_i}$}, where the product involves all the qubits
of the register except the one at site $d$, and let us take the
state with $d=0$ $(\ket{\psi_0}=\ket{\Psi_0})$ as the initial one.
Following the work of Amico {\em et al.}\cite{Fazio}, let us
consider the behavior of the overlap \mbox{${\cal{F}}_{d}(t)=
|\langle \Psi_d|\Psi_0(t)\rangle|^2$}, where $|\Psi_0(t)\rangle$
is the time-evolution of the initial state under the action of
Hamiltonian \eq{Hamiltonian 1}. By looking at \fig{figura1}, we
see that the quantum state diffusion grows indefinitely and the
initial state is never recovered.
Similar numerical results can be obtained in the limit
$\lambda \rightarrow 0$ and nonvanishing local
imperfections\cite{Georgeot1,Georgeot2}.

From \fig{figura1} one must conclude that either
it is possible to perform all the computational processes
in times much shorter than $1/\lambda$ (time scale in which
the fidelity of the state is close to 1), or, otherwise
the quantum register described by \eq{Hamiltonian 1} is not
able to store quantum information for times long enough to
process the desired quantum algorithms.

\subsection{The step-phase modulated quantum register}
\label{mod}

To improve the storing capacity of the quantum register
described in the previous subsection, we implement the
scheme of quantum control that we have briefly outlined
in the Introduction. The idea is to associate to the coupling
amplitude a time dependent phase factor so that the modified
Hamiltonian of the register reads
\begin{equation}\label{Hamiltonian2}
\hat{H}(t) \, = \, -\sum_{i} (\lambda+\chi_i)
\left(e^{i\theta(t)}\sigma_{i}^{+}\sigma_{i+1}^{-} + H. c. \right)
\, + \, B\sum_{i}\sigma_{i}^{z} \; .
\end{equation}
As already mentioned, complex coupling amplitudes can be realized
in systems of charged hopping particles, such as electrons or charged ions hopping along
a string of quantum  dots or Cooper pairs tunnelling on Josephson
junctions arrays\cite{Schoen}, when they are subject to external
electromagnetic vector potentials\cite{Peierls,Scalapino}. To
obtain a spatially constant phase in the intersite hopping
amplitudes, one can consider, for instance, a thin solenoid placed
at the center of a closed, circular ring. Then, in the associated
spin formalism, the phase amplitude $\theta$ of the spin-spin
couplings is equal for any pair of neighboring sites and is
proportional to the ratio of the linked magnetic flux $\phi$ and
the total number $N$ of sites: $\theta \propto
\phi/N$\cite{Aharonovbohm}.

Clearly, not any time modulation of the phase factor
can realize the desired perfect state storage.
As we have already mentioned, two requirements must be
satisfied. The first prescription is that the time modulated
Hamiltonian must obey the commutation property at different times:
$[\hat{H}(t),\hat{H}(t')]=0$. In the case of the modulated Hamiltonian
\eq{Hamiltonian2}, this property is verified if
and only if $\theta(t)-\theta(t')=k \pi$, with $k$ integer.
This implies that we must modulate the phase in such a way that,
during the entire evolution, it regularly jumps on and off
between the two constant values $\theta_0$ and $\theta_0+\pi$,
namely, we must realize a step-phase modulation.

The second property that must be verified in order to obtain a perfect,
time-periodic state reconstruction is that, at a certain given time $T$,
all the time integrals appearing in the Dyson series \eq{evoluzione}
for the quantum state evolution must be equal or differ by a trivial
phase factor integer multiple of $2 \pi$. Let us observe that, independently of
the values of $\theta$, the local term $B\sum_i \sigma_i^z$ commutes
with the $XY$ residual interaction terms in the Hamiltonian \eq{Hamiltonian2}.
Hence, there exists a complete set of
eigenstates of the total Hamiltonian that are as well simultaneous
eigenstates of both the local and the interaction terms. But then,
for all eigenvalues
$\varepsilon^\alpha(\theta_0)$ associated to the eigenstates
$\ket{\alpha}$, we have that:
$\varepsilon^\alpha(\theta_0)=\varepsilon^\alpha_{c}
(\theta_0)+\varepsilon^\alpha_{l}$, where
$\varepsilon^\alpha_{c}(\theta_0)$ and $\varepsilon^\alpha_{l}$
are, respectively, the interaction and the local contributions to
the energy. On the other hand, when $\theta$ passes from
the value $\theta_0$ to $\theta_0+\pi$,
the coupling contribution to the energy changes sign while the
local one remains unchanged, so that
$\varepsilon^\alpha(\theta_0+\pi)=-\varepsilon^\alpha_{c}
(\theta)+\varepsilon^\alpha_{l}$. Therefore any two energy
eigenvalues $\varepsilon^{\alpha}(\theta)$ and
$\varepsilon^{\alpha}(\theta+\pi)$ corresponding to the same
eigenstate $\ket{\alpha}$, differ only in the sign of the
interaction component.

Collecting these facts together, if we select a regular step time modulation of
the phase of the form\cite{precedente}
\begin{equation}\label{modulazione}
\theta(t) \, = \, \left\{
\begin{array}{lll}
\theta & \; \; \; &  0 \leq t < T/2 \\
\theta + \pi & \; \; \; &  T/2 \leq t < T \\
\end{array}
\right.
\end{equation}
periodically repeated for any $t \geq T$, we obtain that the
contribution of the residual interaction Hamiltonian
to the quantum state time evolution vanishes at any time
$t=mT$ with $m$ arbitrary integer. Consequently, it turns out
that by a proper time modulation of the external electromagnetic
potential and of the corresponding Aharonov-Bohm phase, it is possible to
whipe out the effects of the undesired $XY$ couplings in a quantum register.

Regarding the role of the local term, we must discriminate
between two different situations. The good case is when
we can control and choose the period $T$ in such a way that
$BT=2l\pi$ with $l$ integer.
Under this hypothesis {\it any} initial quantum state is reconstructed
{\it exactly} at all times $t$ integer multiples of $T$.
The bad instance is when the parameters are fixed such that
$BT \neq 2l\pi$. If we store an arbitrary initial state
linear combination of states with different numbers of up and down
spins, i.e. states of different magnetizations,
at all times integer multiples of $T$ each of these states
is reconstructed, but, unfortunately, with extra, geometric phase factors
between them, and perfect quantum state reconstruction becomes impossible.
However, even in this case, perfect storage is still achieved in the
subspace of states that can be expressed as linear combinations of
local states all with the same, fixed value of the magnetization.

\begin{figure}[t!]
\includegraphics[width=7.5cm]{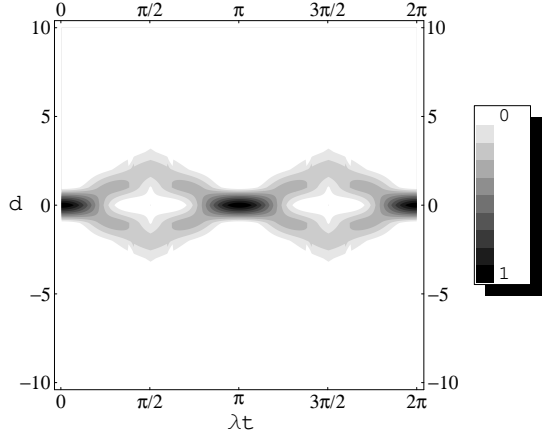}
\caption{Step-periodic time-modulation of the phase,
\eq{modulazione}: two-dimensional contour plot showing the evolution of
the overlap ${\cal{F}}_{d}(t)$, for the same initial state considered
in \fig{figura1}, as a function of the distance $d$
from site $0$ ($y$-axis) and of the dimensionless time $\lambda t$
in atomic units $\hbar=1$ ($x$-axis). Here $\lambda T =\pi$ and
$\theta=\pi/2$. The value of the overlap increases from white
(${\cal{F}}_{d}(t) = 0$) to black (${\cal{F}}_{d}(t) = 1$).}
\label{figura2}
\end{figure}
To compare the unmodulated and the step-phase modulated quantum
registers, in \fig{figura2} we again plot the time evolution of
the overlap ${\cal{F}}_{d}(t)$, for the same initial state considered
in \fig{figura1}. At striking variance with the
unmodulated case reported in \fig{figura1}, we see that the
step-phase modulated register realizes exact, time-periodic
coherent revivals of the initial state. Moreover, we see
from \fig{figura2} that the overall spatial diffusion of the
state is confined in a well defined and narrow region of the
ring. this result can be generalized to
any initial state $\ket{\Psi_0}$ with an arbitrary number of
flipped spins (many magnons). In particular, one finds the same coherent
time-periodic revival of the state as in the one-magnon case,
while the spatial spread becomes a function of the size of the
region along which the initial state is extended, but remains in
any case finite and limited.

\subsection{Possible sources of noise}
\label{err}

Tackling the issue of practical implementations for the quantum
state storage scheme that we have introduced, it is very important
to verify that it does does not depend critically on a
perfect realization of the step-periodic time modulation of the
phase and on other possible sources of noise and imperfections.
\begin{figure}[h!]
\includegraphics[width=8.cm]{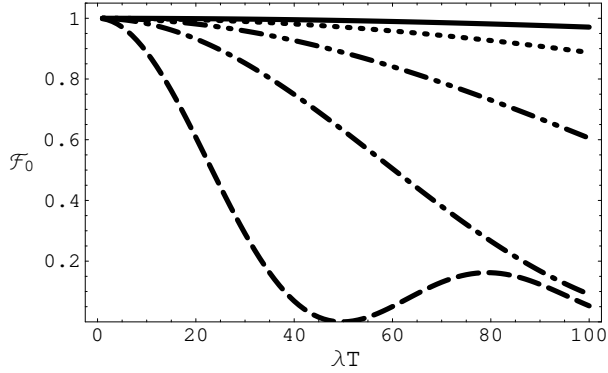}
\caption{Fidelity ${\cal{F}}_{0}(t)$ on site $i=0$ as a function
of the number of periods $\lambda T$ for the initial one-magnon
state $\ket{\Psi_0}$ when we replace the step-periodic
time-modulated phase $\theta(t)$ with its finite-harmonic Fourier
approximations, in increasing order. Dashed line: first 5
harmonics; dot-dashed line: first 13 harmonics; dot-dot-dashed
line: first 25 harmonics; dotted line: first 50 harmonics; solid
line: first 100 harmonics.} \label{figura3}
\end{figure}
We begin by investigating whether the step phase modulated
register can be efficiently implemented when
considering real step function generators of finite precision.
In \fig{figura3} we show the evolution of the fidelity
${\cal{F}}_{0}(t) = |\langle \Psi_0|\psi(t)\rangle|^2$
for the same initial one-magnon state $\ket{\Psi_0}$ as before,
as a function of the number of periods,
when the step-periodic, time-modulated phase
$\theta(t)$ \eq{modulazione} is approximated by its Fourier
decompositions, truncated at various finite orders.
Rearkably, we see that even when considering only the first $100$
harmonics, the fidelity remains close to the ideal limit
${\cal{F}}_{0}(t)=1$ for very long times. This demonstrates
the stability of the finite-harmonic approximation and,
as a consequence, the robustness of our storing protocol
against imperfections in the external control of the phase.

Another important source of noise that modifies the quantum
register described by \eq{Hamiltonian2} is the presence of
local, site-dependent random imperfectins $\eta-i$ in the half-energy
gap, so that the Hamiltonian \eq{Hamiltonian2} is modified and
becomes
\begin{equation}\label{Hamiltonian3}
\hat{H}(t) \, = \, -\sum_{i}(\lambda + \chi_i)
\left(e^{i\theta(t)}\sigma_{i}^{+}\sigma_{i+1}^{-} + H. c. \right)
\, + \, \sum_{i}(B+ \eta_i)\sigma_{i}^{z} \; ,
\end{equation}
where $\eta_i$ are random variables of vanishing mean and
finite variance $\sigma_{\eta}^2$. Because of the presence of the
local noise term $\hat{H}_{err} = \sum_i\eta_i \sigma_i^z$, it is no
longer possible to satisfy the commutation property for any value of
the phase $\theta$, and hence state reconstruction will
not be perfectly achieved, in general.
However it is possible, and important, to determine to what
extent the presence of the local noise affects the
efficiency of the storing scheme. To this aim, we first note
that both the local noise and the
interaction terms commute with the local term
$\hat{H}_{l} = B\sum_i \sigma_i^z$, even if
they do not commute with each other. It is then possible to introduce
the complete set of states $\{\ket{\gamma}\}$ that are
simultaneous eigenstates of the local term, with eigenvalue
$\varepsilon^\gamma_{l}$, and of
\mbox{$\hat{H}_{1}=-\hat{H}_{i} +\hat{H}_{err}$}, with eigenvalue
$\varepsilon^\gamma_{1}$, where $\hat{H}_{i}$ is the $XY$ interaction
in the Hamiltonian \eq{Hamiltonian3}.
After a generic period $T$, perfect reconstruction of the
initial state, written as a linear combination of states
$\{\ket{\gamma}\}$, is assured if the fidelity $\bk{\gamma}{\gamma(T)} =1,
\, \forall \; \ket{\gamma}$.
We can then quantify the effect of the local
noise term by studying the difference of the
fidelities $\bk{\gamma}{\gamma(T)}$ with and without the local
noise term (in the latter case $\bk{\gamma}{\gamma(T)} = 1$ trivially).
With the notations introduced above, and assuming
a step time-modulation of the phase $\theta$,
the application of the total Hamiltonian
\eq{Hamiltonian3} to a generic state $\ket{\gamma}$ yields
\begin{equation}\label{azione}
\hat{H}(t)\ket{\gamma} \, = \, \left\{
\begin{array}{lll}
\left(\varepsilon^\gamma_{l}-\varepsilon^\gamma_{1} \right)
\ket{\gamma}
\, + \, 2 \hat{H}_{err} \ket{\gamma}& \; \; \; &  0 \leq t < T/2 \; ,\\
\left(\varepsilon^\gamma_{l}+\varepsilon^\gamma_{1} \right) \ket{\gamma}
& \;\; \; &  T/2 \leq t < T \; . \\
\end{array}
\right.
\end{equation}
Because the states $\ket{\gamma}$ are not eigenstates of
$\hat{H}_{err}$, we have that
\mbox{$\bra{\gamma'}\hat{H}_{err}\ket{\gamma} \neq0$} for any pair
($\gamma, \; \gamma'$). On the other hand, in the absence of the phase
modulation (unmodulated register with noise), one has
\begin{equation}\label{azione1}
\hat{H}(t)\ket{\gamma} \, = \, \left(\varepsilon^\gamma_{l}-\varepsilon^\gamma_{1}
\right) \ket{\gamma} \, + \, 2 \hat{H}_{err} \ket{\gamma} \; \; \; \forall \; t \; .
\end{equation}
We may now use the expressions in \eq{azione} and \eq{azione1}
to evaluate the fidelity  in the two
different situations by the Dyson series. Assuming for instance a Gaussian
distribution for the local noise $\eta_i$, we obtain,
for the periodic step time modulation and for the case of constant phase,
respectively,
\begin{eqnarray}\label{fidelity}
\bk{\gamma}{\gamma(T)}& \; = \; & \exp{\{ -i
\varepsilon^\gamma_{l}T\}}A(\sigma_\eta,\varepsilon^\gamma_{1},\sigma_{1}) \; ,
\\
 \bk{\gamma}{\gamma(T)}& \; = \; & \exp{\{ -i
(\varepsilon^\gamma_{l}+\varepsilon^\gamma_{1})T\}}A'(\sigma_\eta,\varepsilon^\gamma_{1},\sigma_{1})
\; . \nonumber
\end{eqnarray}
Looking at the exponential terms we immediately note that the
periodic step time modulation of the phase is still effective
in suppressing the undesired action of the residual $XY$ interaction
terms. Concerning the
attenuations terms $A$ and $A'$ in \eqs{fidelity}, both
functions depend on the Gaussian width $\sigma_\eta$
of the local noise term, on the eigenvalue $\varepsilon^\gamma_{1}$
of the Hamiltonian $H_1$, associated to the $\ket{\gamma}$, and on the
spread $\sigma_1$ of the density of states of $H_1$ as a
function of the energy. Both functions must reduce to unity
in the limit of vanishing $\sigma_\eta$ but, in spite of this information,
it is difficult to obtain an analytic expression for any one of them.
However, close to the limit of vaishing $\sigma_\eta$ it is possible
to compare their series expansions, finding that
$A$ remains always closer to one than $A'$. This implies that the
the scheme based on the phase step modulation \eq{modulazione}
of the off-diagonal Hamiltonian terms has an effect on the diagonal (local) part as well,
allowing for a better quantum state storage compared to the unmodulated
register.

\section{Summary and outlook}
\label{conclusion}

We have presented a novel scheme for the storage of
quantum information in a quantum register based on periodic,
perfect state reconstruction. As an example of application of this storing
scheme we have discussed a quantum register based on time
modulation of a Aharonov-Bohm phase. This kind of quantum register
is able to fully cancel the effects of residual interactions of the $XY$
type. Moreover, the scheme appears to be robust even in the presence of
other sources of static noise, such as phase modulations of finite precision
and local noise on the internal Hamiltonian. The study of the effects of
dynamic imperfections is currently under way, and we hope to soon report
on it\cite{futuro}.

\section*{Acknowledgements}

We thank Simone Montangero for interesting discussions.
Financial support from INFM, INFN and MIUR is acknowledged.


\begin{thebibliography}{99}

\bibitem{Bose} S. Bose, Phys. Rev. Lett. {\bf 91}, 207901 (2003).

\bibitem{BoseKorepin} S. Bose, B.-Q. Jin, and V. E. Korepin, quant-ph/0409134 (2004).

\bibitem{Depasquale} F. de Pasquale, G. Giorgi, and S. Paganelli,
Phys. Rev. Lett. {\bf 93}, 120502 (2004).

\bibitem{Subramanyam} V. Subrahmanyam, Phys. Rev. A {\bf 69}, 034304 (2004).

\bibitem{Datta}  M. Christandl, N. Datta, A. Ekert, and A. J. Landahl,
Phys. Rev. Lett. {\bf 92}, 187902 (2004).

\bibitem{Song} Z. Song and C. P. Sun, quant-ph/0412183 (2004), and references therein.

\bibitem{Sun} C. P. Sun, Y. Li, and X. F. Liu, Phys. Rev. Lett. {\bf 91}, 147903 (2003).

\bibitem{Lukin} M. D. Lukin, Rev. Mod. Phys. {\bf 75}, 457 (2003).

\bibitem{Lukin2} M. Fleischhauer and M. D. Lukin, Phys. Rev. Lett. {\bf 84}, 5094
(2000).

\bibitem{Lukin3} M. Fleischhauer and M. D. Lukin, Phys. Rev. A {\bf 65},
022314 (2002).

\bibitem{Lukin4} J. M. Taylor, C. M. Marcus, and M. D. Lukin, Phys. Rev. Lett.
{\bf 90}, 206803 (2003).

\bibitem{nano1} E. Pazy, I. D'Amico, P. Zanardi, and F. Rossi, Phys. Rev. B {\bf 64},
195320 (2001).

\bibitem{nano2} Y. Li, P. Zhang, P. Zanardi, and C. P. Sun, Phys. Rev. A {\bf 70},
032330 (2004).

\bibitem{Zoller} A. Imamoglu, E. Knill, L. Tian, and P. Zoller, Phys. Rev. Lett.
{\bf 91}, 017402 (2003).

\bibitem{Poggio} M. Poggio, G. M. Steeves, R. C. Myers, Y. Kato, A. C. Gossard, and D. D. Awschalom,
Phys. Rev. Lett. {\bf 91}, 207602 (2003).

\bibitem{Gingrich}  R. M. Gingrich, P. Kok, H. Lee, F. Vatan, and J. P. Dowling,
Phys. Rev. Lett. {\bf 91}, 217901 (2003).

\bibitem{Kitaev} E. Dennis, A. Kitaev, A. Landahl, and J. Preskill, J. Math. Phys. {\bf 43}, 4452 (2002).

\bibitem{precedente} S. M. Giampaolo, F. Illuminati, A. Di Lisi, and S. De Siena, quant-ph/0503107 (2005).

\bibitem{Georgeot1} B. Georgeot and D. L. Shepelyansky, Phys. Rev. E {\bf 62}, 3504 (2000).

\bibitem{Georgeot2} B. Georgeot and D. L. Shepelyansky, Phys. Rev. E {\bf 62}, 6366 (2000).

\bibitem{Montangero}  G. De Chiara, D. Rossini, S. Montangero, and
R. Fazio, \mbox{quant-ph/0502148}.

\bibitem{Schoen} Y. Makhlin, G. Sch\"on and A. Shnirman, Rev. Mod. Phys. {\bf 73}, 357
(2001).

\bibitem{Feynman57} R. P. Feynman, F. L. Vernon, and R. W. Hellwarth,
J. Appl. Phys. {\bf 28}, 49 (1957).

\bibitem{Peierls} R. E. Peierls, Z. Phys. {\bf 80}, 763 (1933).

\bibitem{Aharonovbohm} Y. Aharonov and D. Bohm, Phys. Rev. {\bf
115}, 485 (1959).

\bibitem{Scalapino} D. J. Scalapino, S. R. White, and S. Zhang,
Phys. Rev. B {\bf 47}, 7995 (1993).

\bibitem{Fazio} L. Amico, A. Osterloh, F. Plastina, R. Fazio,
and G. M. Palma, Phys. Rev. A {\bf 69}, 022304 (2004).

\bibitem{futuro} S. M. Giampaolo, F. Illuminati, and S. Montangero, in preparation.

\end{thebibliography}
\end{document}